\documentclass[]{spie}  %>>> use this instead for A4 paper
\usepackage{amsmath,amsfonts,amssymb}
\usepackage{graphicx}
\usepackage[colorlinks=true, allcolors=blue]{hyperref}

\title{A compact and modular X and gamma-ray detector with a CsI scintillator and double-readout Silicon Drift Detectors}

\author[a,b]{R.~Campana}
\author[b,c]{F.~Fuschino}
\author[a,b]{C.~Labanti}
\author[a]{M.~Marisaldi}
\author[a]{L.~Amati}
\author[d]{M.~Fiorini}
\author[d]{M.~Uslenghi}
\author[b,c]{G.~Baldazzi}
\author[e]{P.~Bellutti}
\author[f, g]{Y.~Evangelista}
\author[h]{I.~Elmi}
\author[f, g]{M.~Feroci}
\author[e]{F.~Ficorella}
\author[a,i]{F.~Frontera}
\author[e]{A.~Picciotto}
\author[e]{C.~Piemonte}
\author[l]{A.~Rachevski}
\author[l,m]{I.~Rashevskaya}
\author[b,c]{L.~P.~Rignanese}
\author[l,n]{A.~Vacchi}
\author[l]{G.~Zampa}
\author[l]{N.~Zampa}
\author[e]{N.~Zorzi}
\affil[a]{INAF/IASF-Bologna, Via Gobetti 101, I-40129 Bologna, Italy}
\affil[b]{INFN - Sezione di Bologna, Viale Berti Pichat 6/2, I-40127 Bologna, Italy}
\affil[c]{Department of Physics, University of Bologna, Viale Berti Pichat 6/2, I-40127 Bologna, Italy}
\affil[d]{INAF/IASF-Milano, Via Bassini 23, I-20133 Milano, Italy}
\affil[e]{FBK - Fondazione Bruno Kessler, Via S. Croce 77, I-38122, Trento, Italy}
\affil[f]{INAF/IAPS, Via Fosso del Cavaliere 100, I-00133 Roma, Italy}
\affil[g]{INFN - Sezione di Roma Tor Vergata, Via della Ricerca Scientifica 1, I-00133 Roma, Italy}
\affil[h]{CNR/IMM, Via Gobetti 101, I-40129 Bologna, Italy}
\affil[i]{Department of Physics and Earth Sciences, University of Ferrara, Via Giuseppe Saragat 1, I-44122 Ferrara, Italy}
\affil[l]{INFN - Sezione di Trieste, Padriciano 99, I-34127 Trieste, Italy}
\affil[m]{INFN/TIFPA, Via Sommarive 14, I-38123 Povo, Trento, Italy}
\affil[n]{Department of Mathematics, Computer Science and Physics, University of Udine, I-33100 Udine, Italy}

\authorinfo{Further author information: (Send correspondence to R. Campana)\\E-mail: campana@iasfbo.inaf.it}

\begin{document} 
\maketitle

\begin{abstract}
A future compact and modular X and gamma-ray spectrometer (XGS) has been designed and a series of prototypes have been developed and tested. The experiment envisages the use of CsI scintillator bars read out at both ends by single-cell 25 mm$^2$ Silicon Drift Detectors. Digital algorithms are used to discriminate between events absorbed in the Silicon layer (lower energy X rays) and events absorbed in the scintillator crystal (higher energy X rays and $\gamma$-rays). The prototype characterization is shown and the modular design for future experiments with possible astrophysical applications (e.g. for the THESEUS mission proposed for the ESA M5 call) are discussed.
\end{abstract}

\keywords{High energy astrophysics -- X-ray detectors -- $\gamma$-ray detectors}

\section{INTRODUCTION}\label{s:intro}  
Gamma-Ray Bursts (GRB) are one of the most intriguing and challenging phenomena for modern science. 
Because of their huge luminosities (up to more than 10$^{52}$ erg/s), 
their redshift distribution extending from $z\sim0.01$ up to $z>9$ (i.e., much above that of SNe Ia and galaxy clusters), 
their association with peculiar core-collapse supernovae and with neutron star / black-hole mergers, 
their study is of very high interest for several fields of astrophysics. 
These include, e.g., the physics of matter in extreme conditions and plasma physics, black hole physics, core-collapse SNe, cosmology and fundamental physics, production of gravitational wave signals. 
Despite the huge observational advances occurred in the last 15 years several open issues still affect our comprehension of these phenomena, and their exploitation for fundamental physics and cosmology. 
Among the most relevant aspects still lacking a complete understanding are all those connected with the so called ``prompt'' emission. 
A better knowledge of the involved emission processes and of the source geometry is essential, not only to clarify the origin of the ``central engine'' and its connection with the progenitors, but also to assess the real energy budget of different classes of GRBs eventually allowing us to use the GRBs for fundamental physics and cosmology studies. 

To address these fundamental issues, time resolved spectroscopy (and possibly polarimetry) of the GRB prompt emission over a broad energy range from $\sim$1--2 keV (i.e. well below the range of past, present and near future GRB detectors) to several MeV is needed.

We are therefore working towards a prototype (\emph{X and Gamma Spectrometer}, XGS) for a monolithic system which would allow detection, spectroscopy and timing of GRBs and other high energy transients, like soft gamma-ray repeaters, over an unprecedented broad energy band. 
It is therefore possible to obtain a detector with unique capabilities, such as the combination of a low energy threshold (1--2 keV) and energy resolution significantly better than that of any other GRB detection system based on scintillators (e.g., Fermi/GBM) or CZT/CdTe semiconductors (INTEGRAL/ISGRI, Swift/BAT), besides timing capabilities down to a few micro-second resolution over the whole energy band, and possible polarimetric capabilities thanks to the true three-dimensional photon interaction reconstruction.

In this paper we report on the progress of this currently undergoing project.

\section{THE XGS INSTRUMENT CONCEPT}

\begin{figure}[htbp]
\centering
\includegraphics[width=0.75\textwidth]{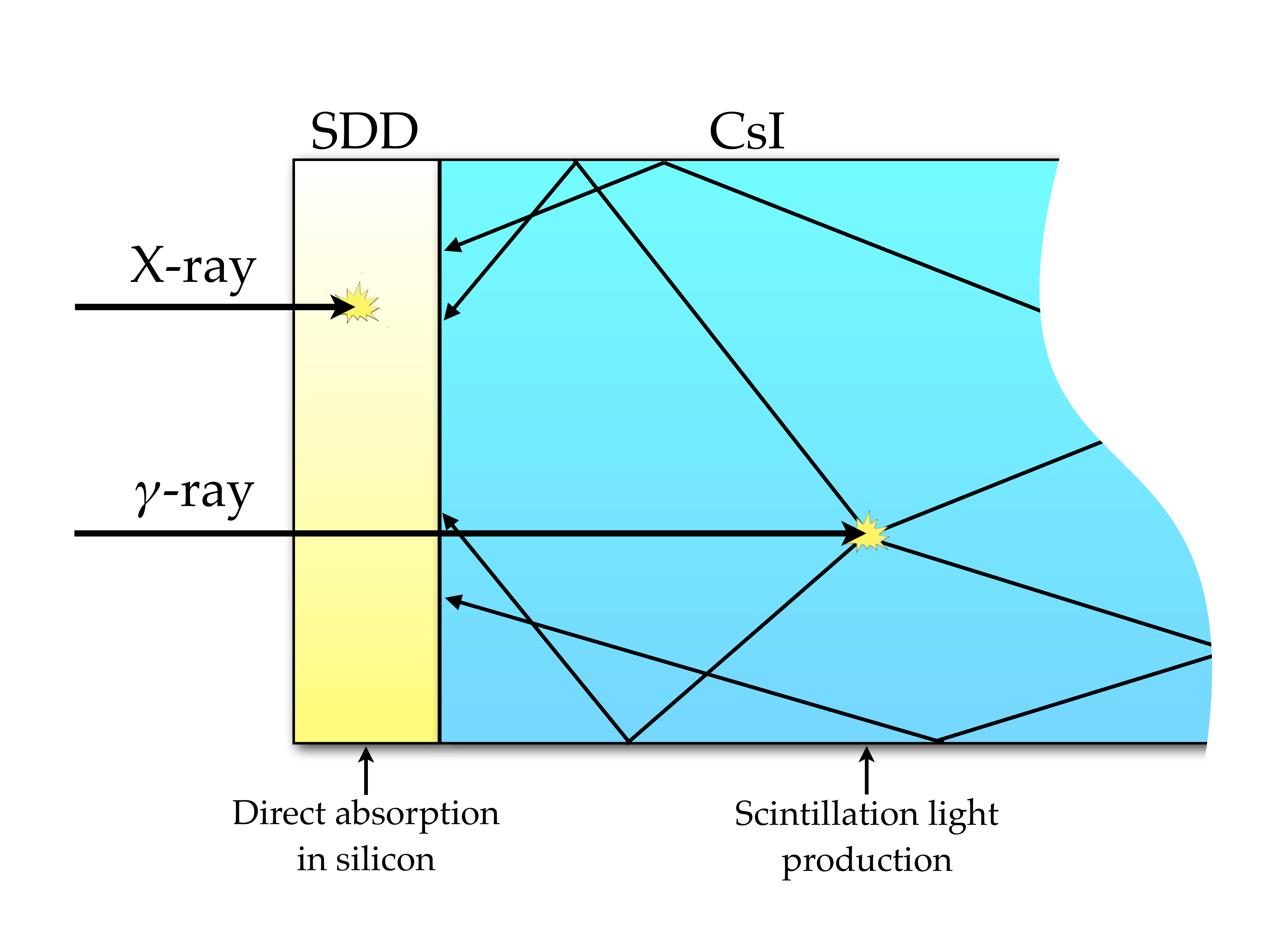}
\caption{The ``siswich'' detector concept. Low energy X-rays are absorbed directly in the silicon bulk of the SDD (X-events), while high energy X-rays and $\gamma$-rays produce scintillation light in the CsI crystal (S-events), that is collected by the SDDs.}
\label{f:siswich_concept}
\end{figure}

Aiming at designing a compact instrument with a very large sensitivity band, the XGS instrument utilizes the so-called ``siswich'' concept \cite{marisaldi04,marisaldi05}, exploiting the coupling between Silicon detectors ($\sim$2 keV up to 30--50 keV) and inorganic scintillator bars (20 keV up to several MeV).

In this concept (Figure~\ref{f:siswich_concept}), the Silicon detectors on the ``top'' side of the bars play the double role of read-out of the signal from the scintillator and of independent detectors. 
The inorganic scintillator bar, made of thallium-doped caesium iodide, is coupled at both ends with a Silicon Drift Detector \cite{gatti84}, developed by INFN-Trieste and Fondazione Bruno Kessler (FBK, Trento) within the framework of the ReDSoX collaboration.
Lower energy X-rays are stopped in one Silicon detector, while higher energy X-rays and $\gamma$-rays are absorbed by the crystal and the optical scintillation photons are collected by the two SDDs. 
The two types of events are distinguished by pulse shape discrimination techniques, through the different risetimes of the corresponding preamplifier signal.
In the case of ``X-events'', the risetime is dominated by the anode collection time ($\sim$100 ns), while for ``S-events'' the signal rises following the characteristic CsI(Tl) scintillation time constants and different light paths, amounting to a few $\mu$s.
The sensitivity band is therefore from $\sim$1 to 50 keV (X-events) and from 20 keV to several MeVs (S-events), allowing for some overlap of the two operating modes.

\begin{figure}[htbp]
\centering
\includegraphics[width=\textwidth]{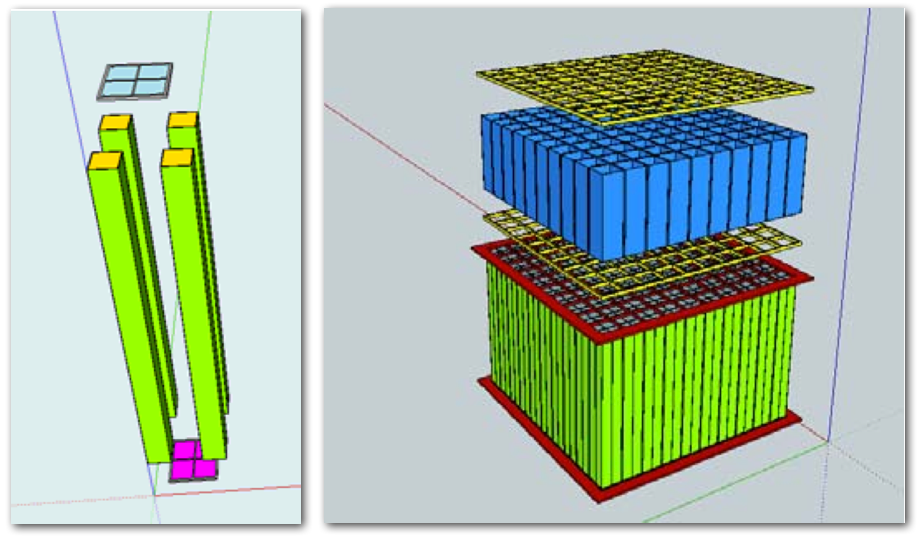}
\caption{\emph{Left panel:} the fundamental unit of XGS, composed of 4 scintillator bars coupled to two 2$\times$2 SDD arrays. \emph{Right panel:} the complete XGS module.}
\label{f:xgs_concept}
\end{figure}

The instrument concept is shown in Figure~\ref{f:xgs_concept}. Four scintillator bars (each about 4.5$\times$4.5$\times$45 mm$^3$) are coupled with an array of 2$\times$2 single-cell SDDs, each with an active area of 25 mm$^2$. This constitutes the fundamental unit of the instrument module, that is composed of, e.g., 64 or 100 bars (Figure~\ref{f:xgs_concept}, right panel) and thus having 128 or 200 electronic channels. To restrict the field of view to about 0.15 sr, and increase the sensitivity of the instrument, a collimator is placed on the top of the bar/detectors assembly.

\section{PRELIMINARY PROTOTYPES}\label{s:prototypes}

\subsection{Aim and front-end electronics}
\begin{figure}[htbp]
\centering
\includegraphics[width=\textwidth]{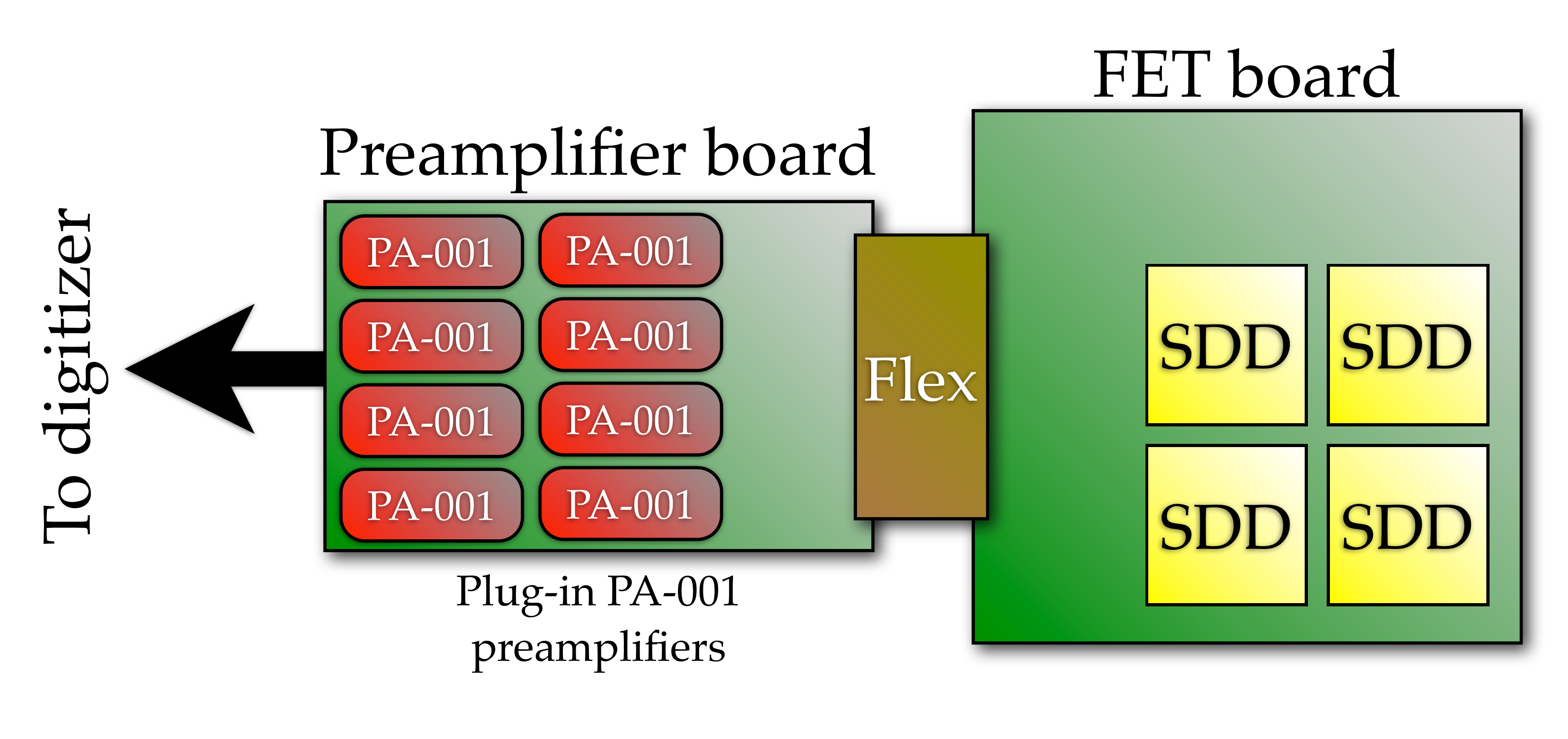}
\caption{The overall setup of the XGS prototype.}
\label{f:overall_setup}
\end{figure}

\begin{figure}[htbp]
\centering
\includegraphics[height=0.35\textwidth]{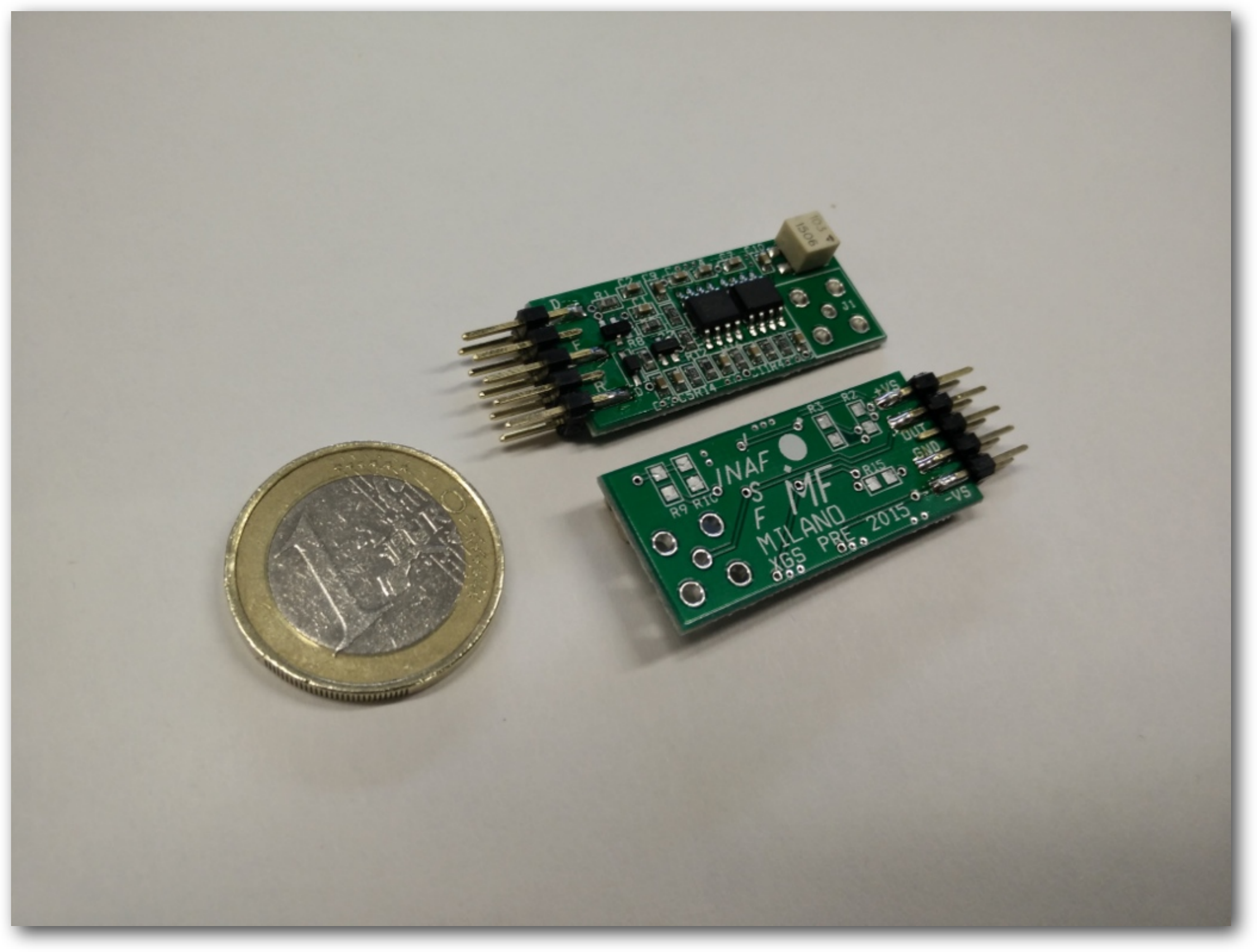}
\includegraphics[height=0.35\textwidth]{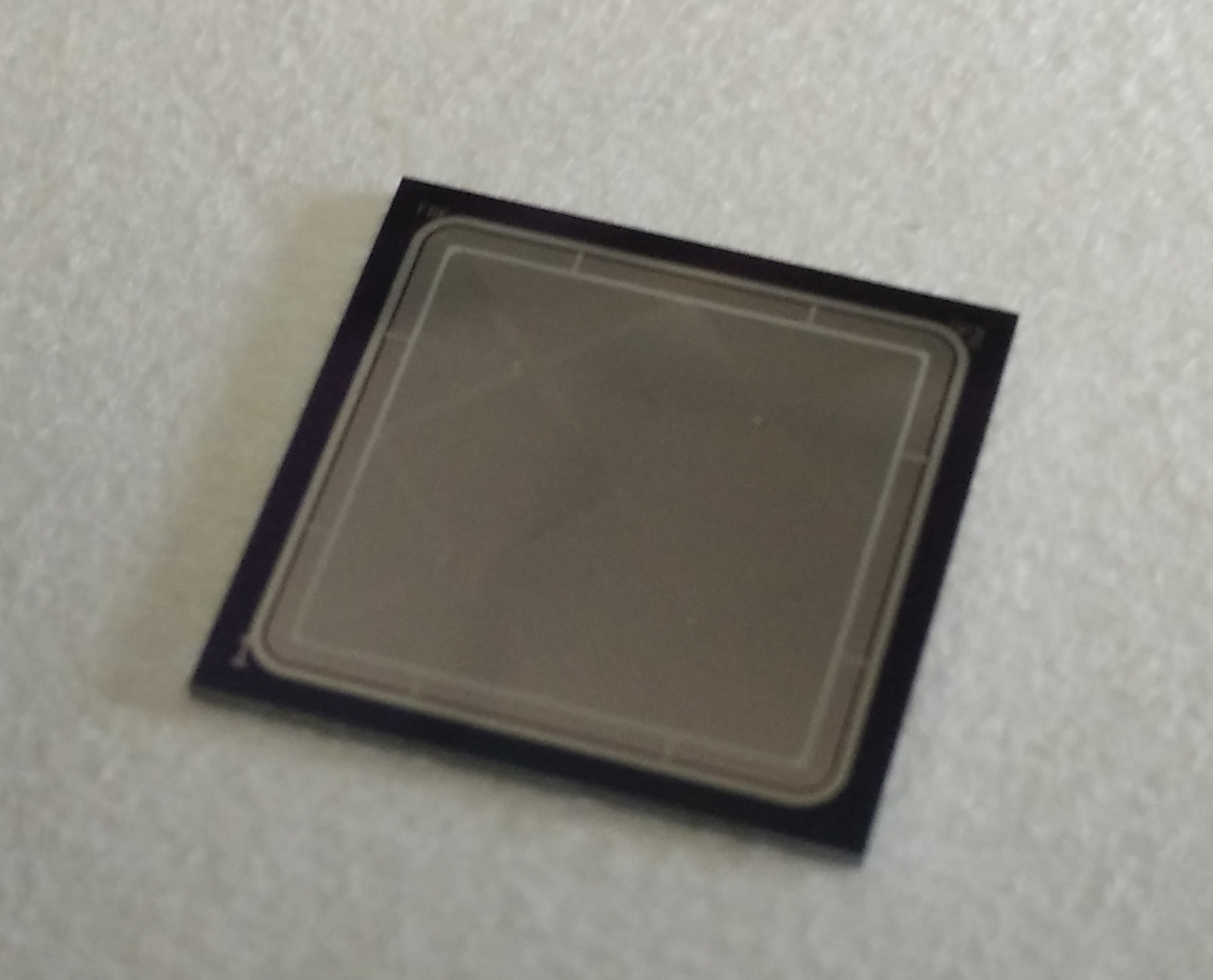}
\caption{\emph{Left panel:} the PA-001 continuous reset charge sensitive preamplifier, in its plug-in configuration. Both sides of the PCB are shown.
\emph{Right panel:} the 2$\times$2 cell SDD array.}
\label{f:pa001}
\end{figure}

In order to demonstrate the proposed instrument architecture, a prototype is being developed in the framework of an INAF-funded project. The overall setup is shown in Figure~\ref{f:overall_setup}. The front-end electronics is composed of a discrete-elements charge sensitive preamplifier (PA-001, developed at INAF/IASF, Figure~\ref{f:pa001}, left panel), with continuous reset of the feedback capacitor. In order to minimize the stray capacitance, the preamplifier first stage input FET is bonded near the anode, in a dedicated PCB board that distributes also the detector bias voltages.
The board houses four 2$\times$2 single-cell SDD arrays (Figure~\ref{f:pa001}, right panel), for a total of 16 channels. The complete detector is composed of four such submodules (16 scintillator bars and 2$\times$16 SDDs, one for each side of the bars).
A flex cable carries the FET output signal to a plug-in motherboard, where single channel preamplifier are connected.
The preamplifier output is then fed to a fast digitizer and then digitally processed.

The choice of a digital signal processing was justified by its large flexibility in this development phase, since it allows simultaneous testing of different shaping algorithms, to characterize the noise performance of the system in a rapid and extremely flexible way.
Of course a custom-designed analog ASIC will be mandatory for an actual space-ready instrument, and the present prototype will allow to drive the design of this front-end electronics component.

\subsection{ProtoXGS-1}
\begin{figure}[htbp]
\centering
\includegraphics[width=0.49\textwidth]{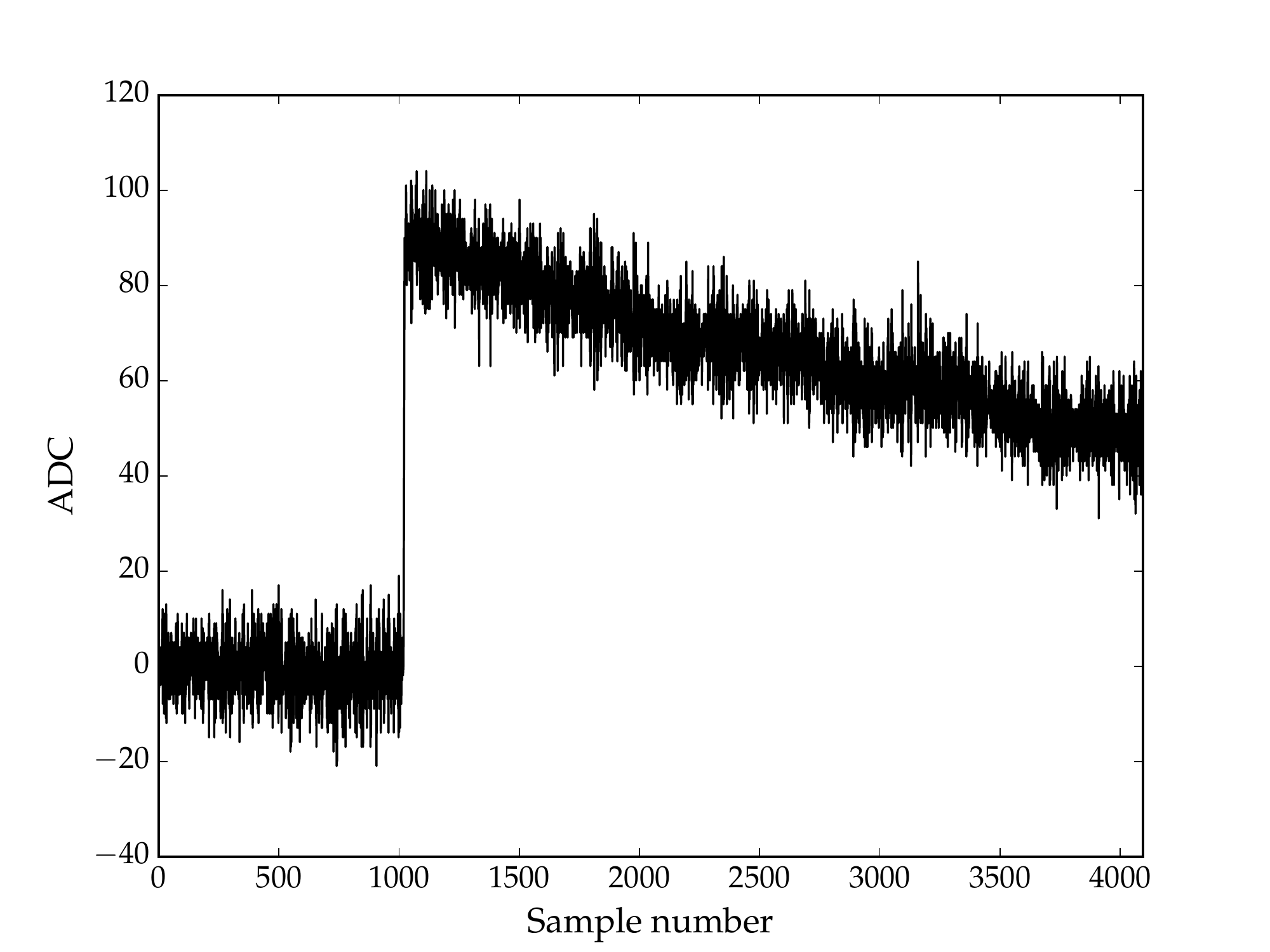}
\includegraphics[width=0.49\textwidth]{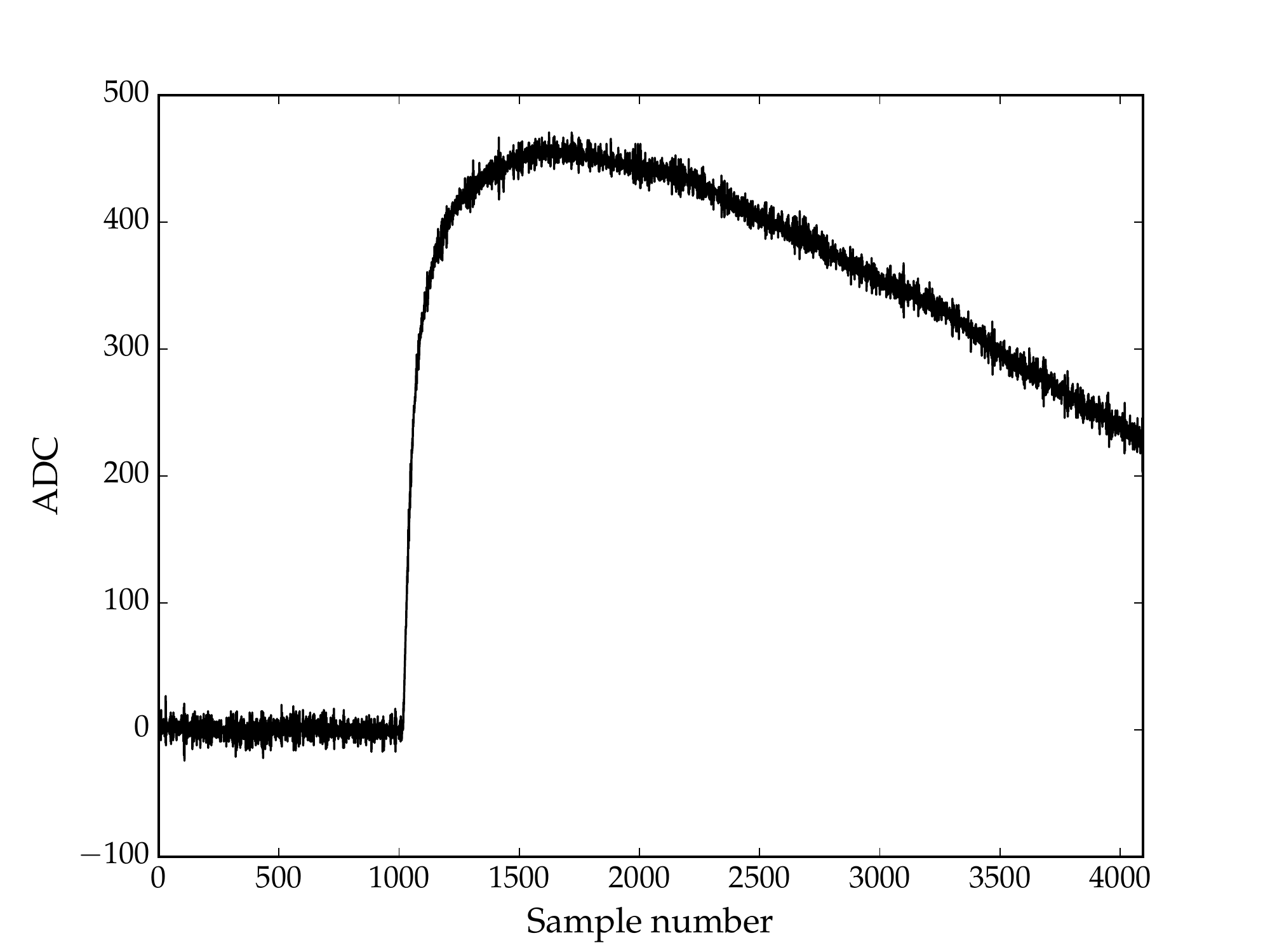}
\caption{\emph{Left panel:} preamplifier output for a 13.94 keV X-event. \emph{Right panel:} preamplifier output for a 662 keV S-event. Both waveforms were digitized with a 30 MS/s sampling rate and a 12 bit ADC.}
\label{f:events}
\end{figure}

A first prototype to test the digital processing algorithms was implemented coupling a 3 mm radius, 1 cm long CsI cylinder with a single-cell 5$\times$5 mm$^2$ SDD. The detector anode is readout by a PA-001 preamplifier and its output acquired using a National Instruments PXI-5105 12-bit, 60 MS/s digitizer. In Figure~\ref{f:events} are shown the two types of events. Typical amplitude of X-ray events directly absorbed by the SDD is about 15 mV for a 13 keV photon, while for 662 keV S-events the amplitude is about 20 mV.

The digitized waveform was then processed with a digital trapezoidal filter \cite{guzik13}. The trapezoidal filter, characterized by symmetric rise and fall times and a flat-top time, is easily implemented as a cascade of finite and infinite impulse response filters, and is a good approximation of the ``ideal'' filter. The trapezoid was then sampled around the middle of the flat-top to extract the pulse height, proportional to the input energy.

\begin{figure}[htbp]
\centering
\includegraphics[width=0.7\textwidth]{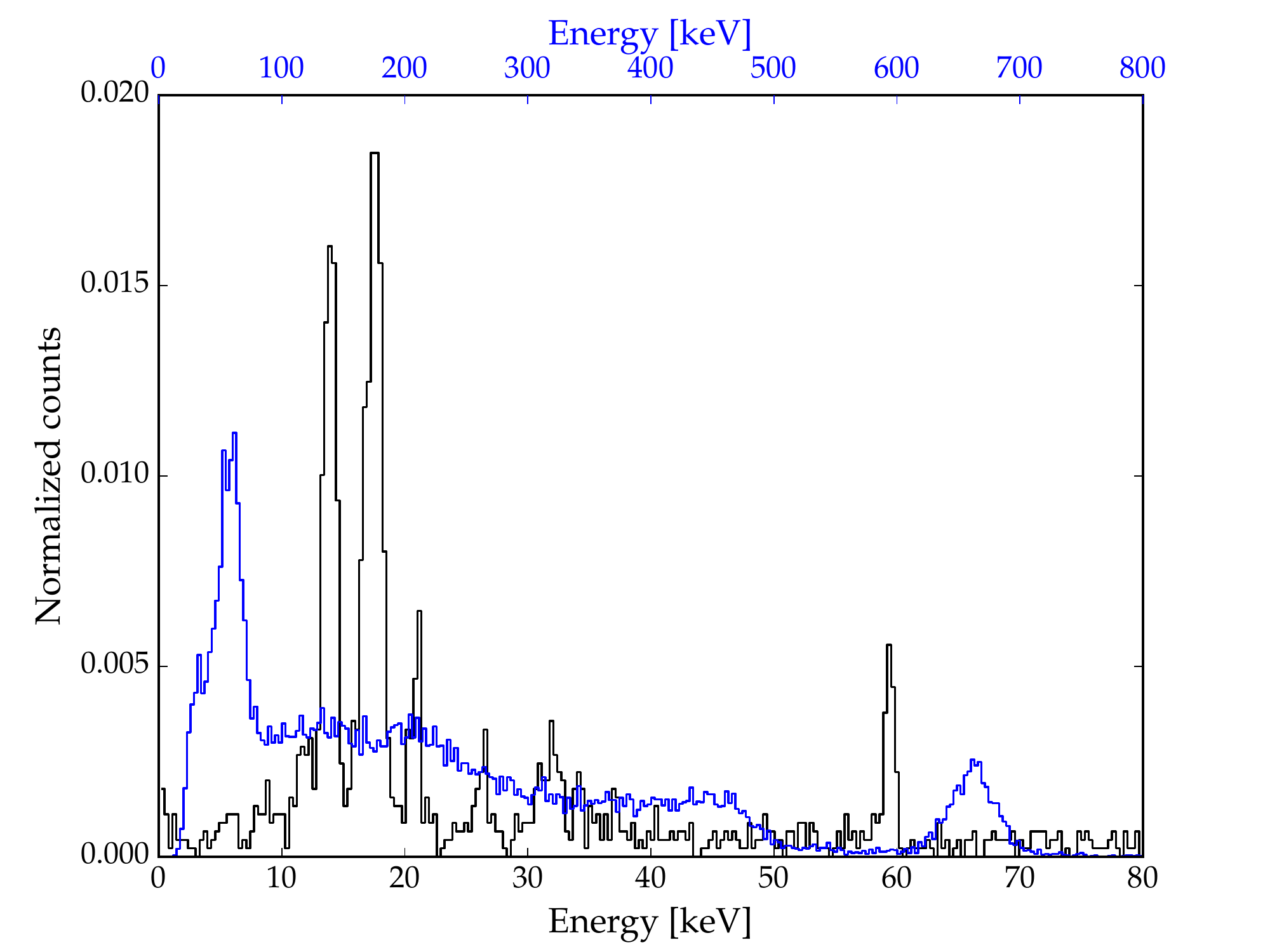}
\caption{Simultaneous acquisition of a $^{137}$Cs plus an $^{241}$Am radioactive source, as seen by ProtoXGS-2. The black curve (lower $x$-axis) is the spectrum of X-events, while the blue one (upper $x$-axis) is the spectrum of S-events. Both types of events were acquired with a digitizer, distinguished with pulse shape discrimination and shaped with a trapezoidal filter.}
\label{f:spectrum}
\end{figure}

In Figure~\ref{f:spectrum} is shown a simultaneous acquisition at room temperature of two radioactive sources, $^{137}$Cs (line at 662 keV plus Compton continuum) and $^{241}$Am (low energy X-ray lines between 11 and 26~keV, and line at 59~keV). 
Both types of events were acquired simultaneously by the digitizer and then distinguished with pulse shape discrimination on the rising edge. 
Optimized digital filters were then applied separately for the two categories.
In this case, a resolution of about 5.8\% on the 662 keV line is obtained.
The good performance of the DSP methods were demonstrated, showing a comparable result with respect to a more traditional acquisition chain composed by an analog shaping amplifier coupled to a commercial multichannel analyzer.

\subsection{ProtoXGS-2}

\begin{figure}[htbp]
\centering
\includegraphics[width=0.7\textwidth]{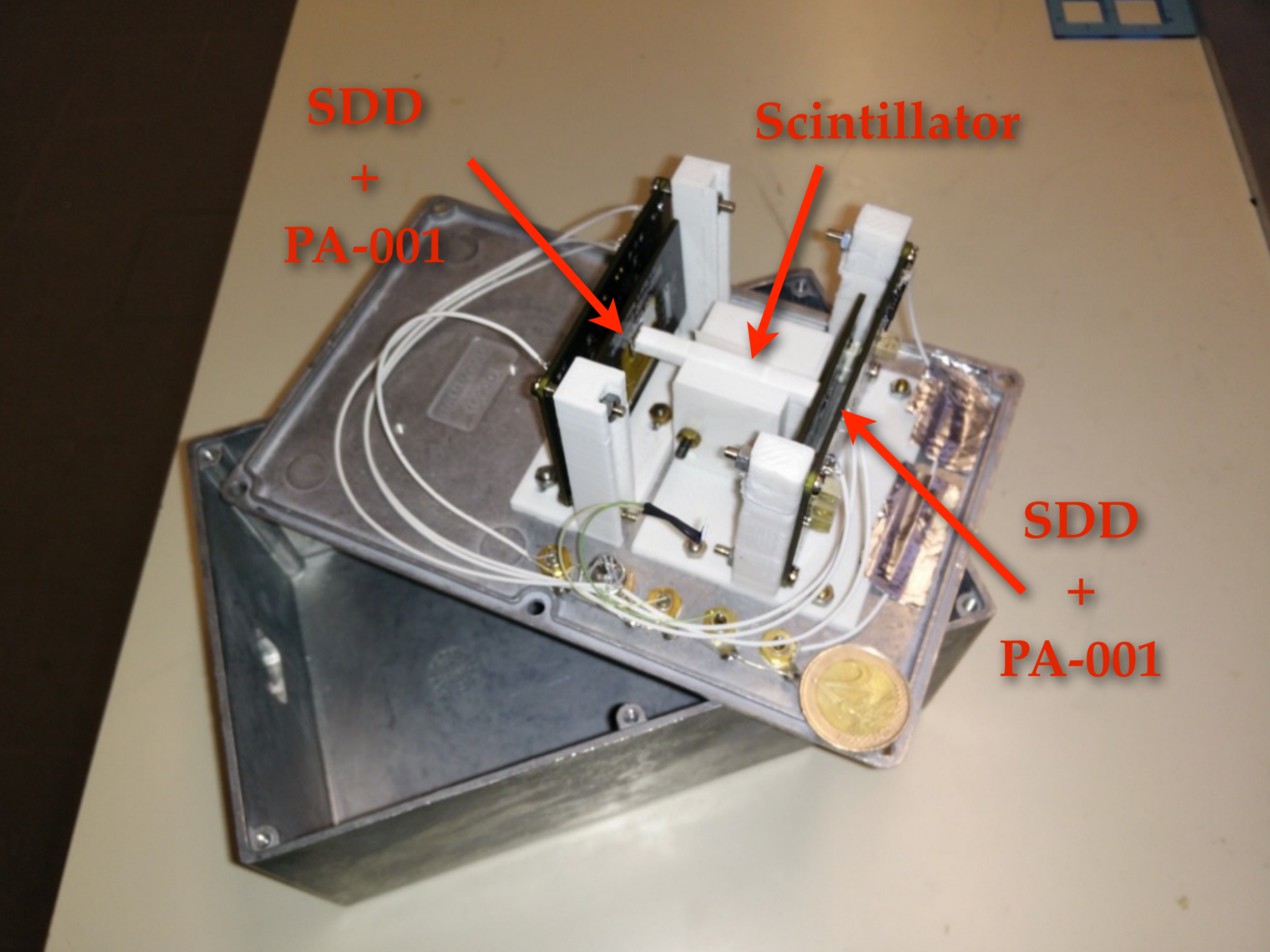}
\caption{The ProtoXGS-2 prototype. A single 4.5$\times$4.5$\times$45 mm$^3$ scintillator bar is coupled at both ends with single cell, 5$\times$5 cm$^2$ SDDs each with a PA-001 preamplifier.}
\label{f:protoxgs2}
\end{figure}

Another prototype (ProtoXGS-2, Figure~\ref{f:protoxgs2}) was developed in order to investigate the optical coupling between the scintillator bar and the SDDs, and to derive realistic performance figures for the complete XGS prototype using a single-bar, 2-channel sub-unit.

The scintillator bar is the same envisaged for XGS, with lengths 4.5$\times$4.5$\times$45 mm$^3$. Also in this case the SDD anodes are coupled to a PA-001 preamplifier whose output is digitized in a similar way with respect to ProtoXGS-1.

Measurements of the light attenuation, spatial and energetic resolution were performed using a collimated ($\sim$2 mm spot) $^{137}$Cs radioactive source at various positions along the bar.
The signal collected at each SDD can be represented, in a simplified model, as \cite{labanti91,labanti08}:
\[
U(x) = E u_0 e^{-\alpha(L/2-x)}
\]
where $L$ is the bar length, $x$ is the position along the bar ($x=0$ is the bar center), $\alpha$ is the light attenuation coefficient, $E$ the input energy and $u_0$ is the signal output at the bar edge (that includes the light yield of the scintillator and the quality of the optical coupling).
The energy and the interaction position can be derived from the signals $U_A$, $U_B$ collected by both detectors as:
\[
E \propto \sqrt{U_AU_B}
\]
and
\[
x \propto \log{\frac{U_A}{U_B}} \, .
\]

\begin{figure}[htbp]
\centering
\includegraphics[width=0.49\textwidth]{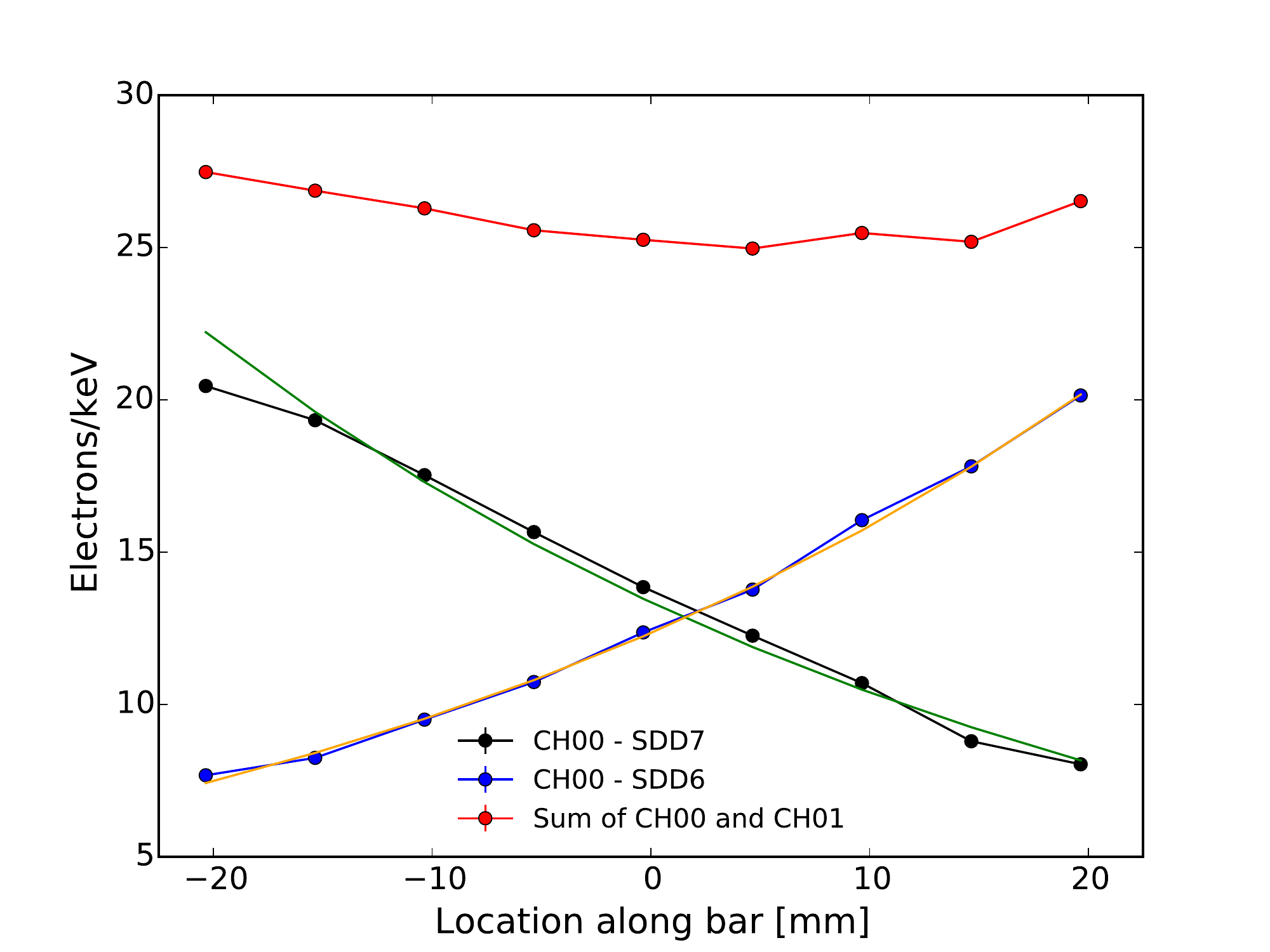}
\includegraphics[width=0.49\textwidth]{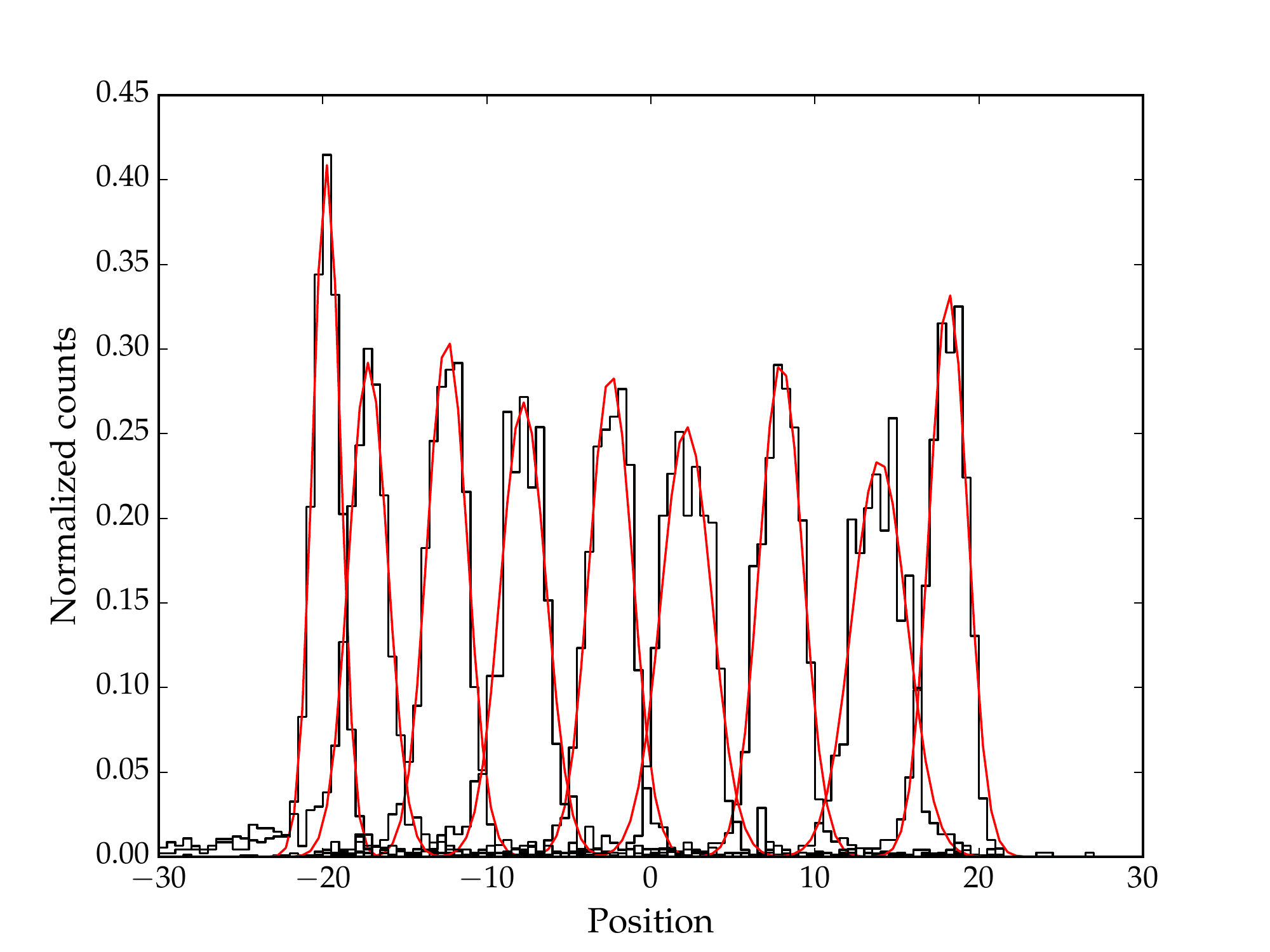}
\caption{\emph{Left panel:} Light attenuation along the bar. The signal from both detectors is shown. \emph{Right panel:} Reconstructed position for various measurements along the bar.}
\label{f:protoxgs2_results}
\end{figure}

In Figure~\ref{f:protoxgs2_results}, left panel, the light attenuation measured by both detectors is shown. The attenuation coefficient is $\sim$0.021 cm$^{-1}$, consistent with previous measurements \cite{marisaldi06, labanti08}. The collective light output is $\sim$27 e$^-$/keV.
In the right panel of Figure~\ref{f:protoxgs2_results} the position reconstruction along the bar is shown. The average FWHM of the reconstructed position is about 3~mm. Considering the source spot size ($\sim$2 mm), the derived intrinsic resolution in position is around 2.2 mm.

\begin{figure}[htbp]
\centering
\includegraphics[width=0.7\textwidth]{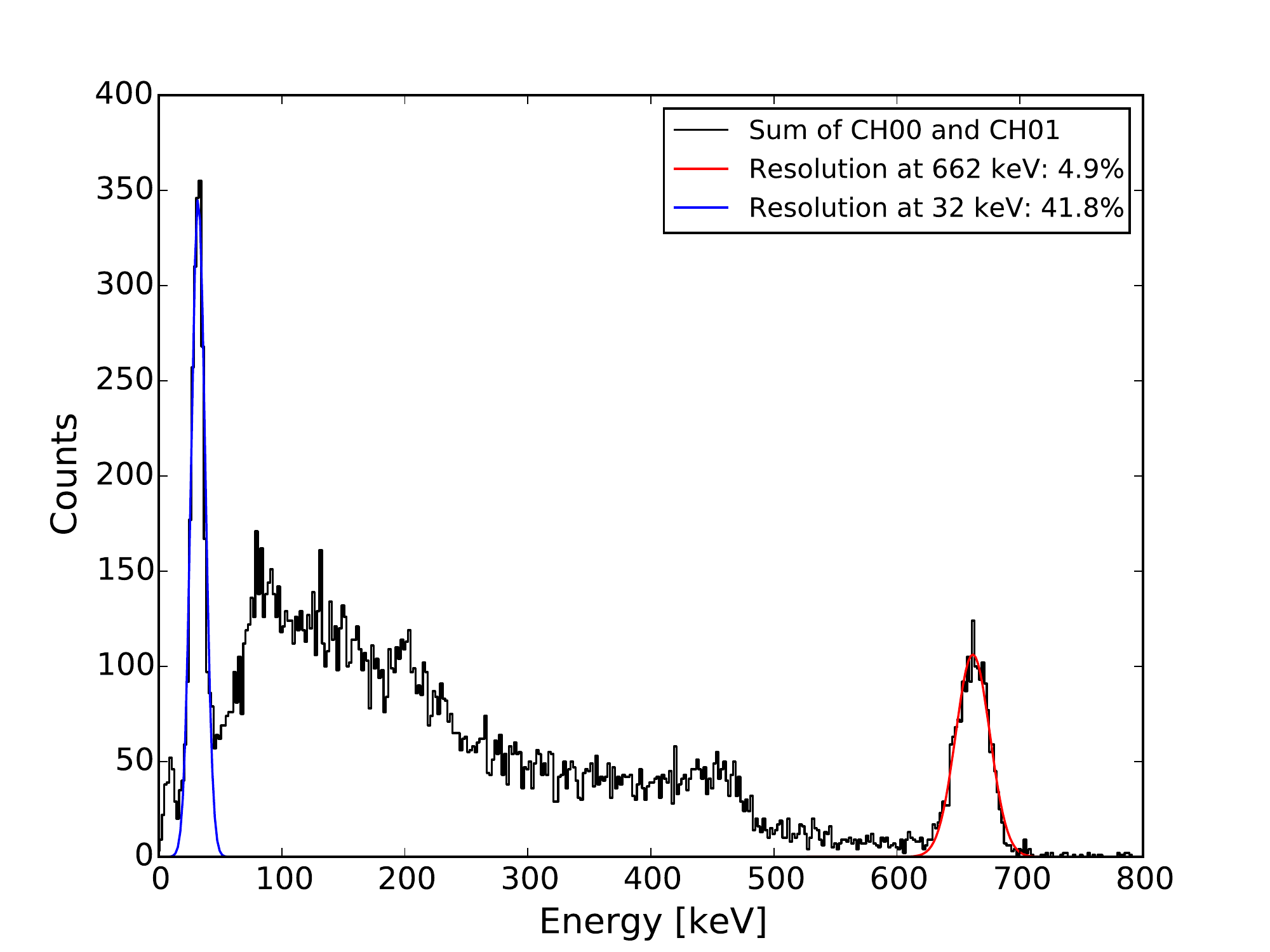}
\caption{Spectrum of a collimated $^{137}$Cs radioactive source, placed in the middle of the scintillator bar, acquired summing the scintillation light signal collected by the SDDs at both ends. Gaussian fits of the 662 and 32 keV peaks are also shown.}
\label{f:protoxgs2_spectrum}
\end{figure}

In Figure~\ref{f:protoxgs2_spectrum} a spectrum obtained with the source at the bar center is shown. The measured resolution at 662~keV is 4.9\%, with a lower threshold of about 20~keV, thus confirming the expected results and the good performance of the system.

\section{BACK-END ARCHITECTURE}\label{s:backend}

\subsection{Digital signal processing with RedPitaya}
For the complete XGS prototype, a digital back-end architecture able to acquire the preamplifier output of several tens of channels, with a suitable trigger logic, is required. The digitized waveforms should be then digitally processed in real-time or offline.

An attractive and low-cost solution envisages the use of a commercial ADC board, the RedPitaya\footnote{\url{www.redpitaya.com}}. This board is based on a Xilinx Zynq 7010 system-on-chip, in which a FPGA is coupled to an ARM Cortex 9 CPU, allowing a great flexibility in the deployment and configuration. Each board houses two analog input channels, with a 14-bit, 125 MS/s ADC, and can be connected to a network via a Gigabit Ethernet port.
A Linux operating system, loaded via a microSD drive, runs on the ARM CPU.
An interesting feature is the availability of the source code for both the operating system and the FPGA core\footnote{\url{https://github.com/RedPitaya/RedPitaya}}, allowing for an easy customization, e.g. by adding desired features to the default configuration.

A first scalable prototype, using two boards and thus able to read out 4 analog channels, has been developed (Figure~\ref{f:pitayas}, left panel).

\subsection{Trigger logic and board synchronization}

\begin{figure}[htbp]
\centering
\includegraphics[width=0.4\textwidth]{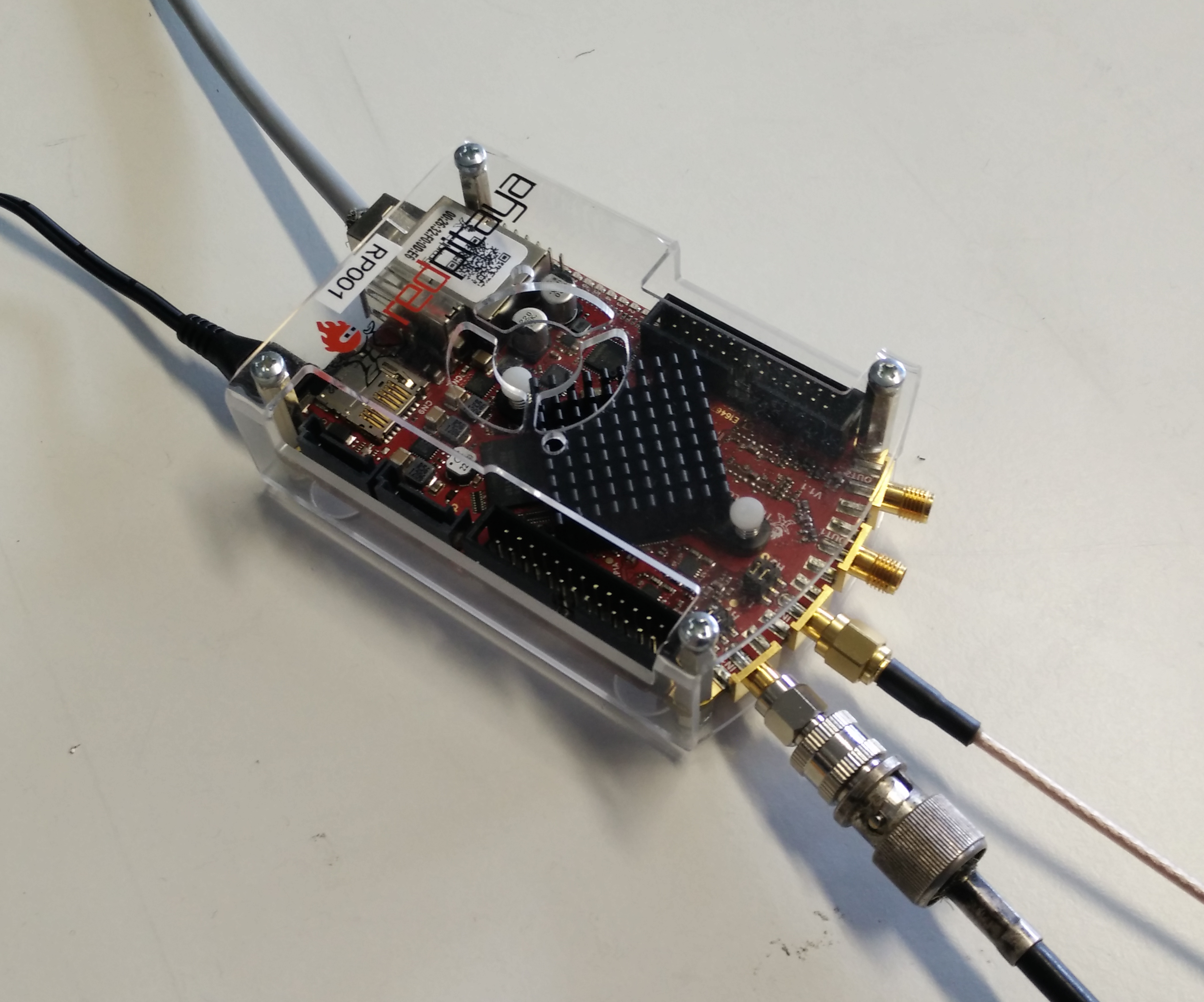}
\includegraphics[width=0.5\textwidth]{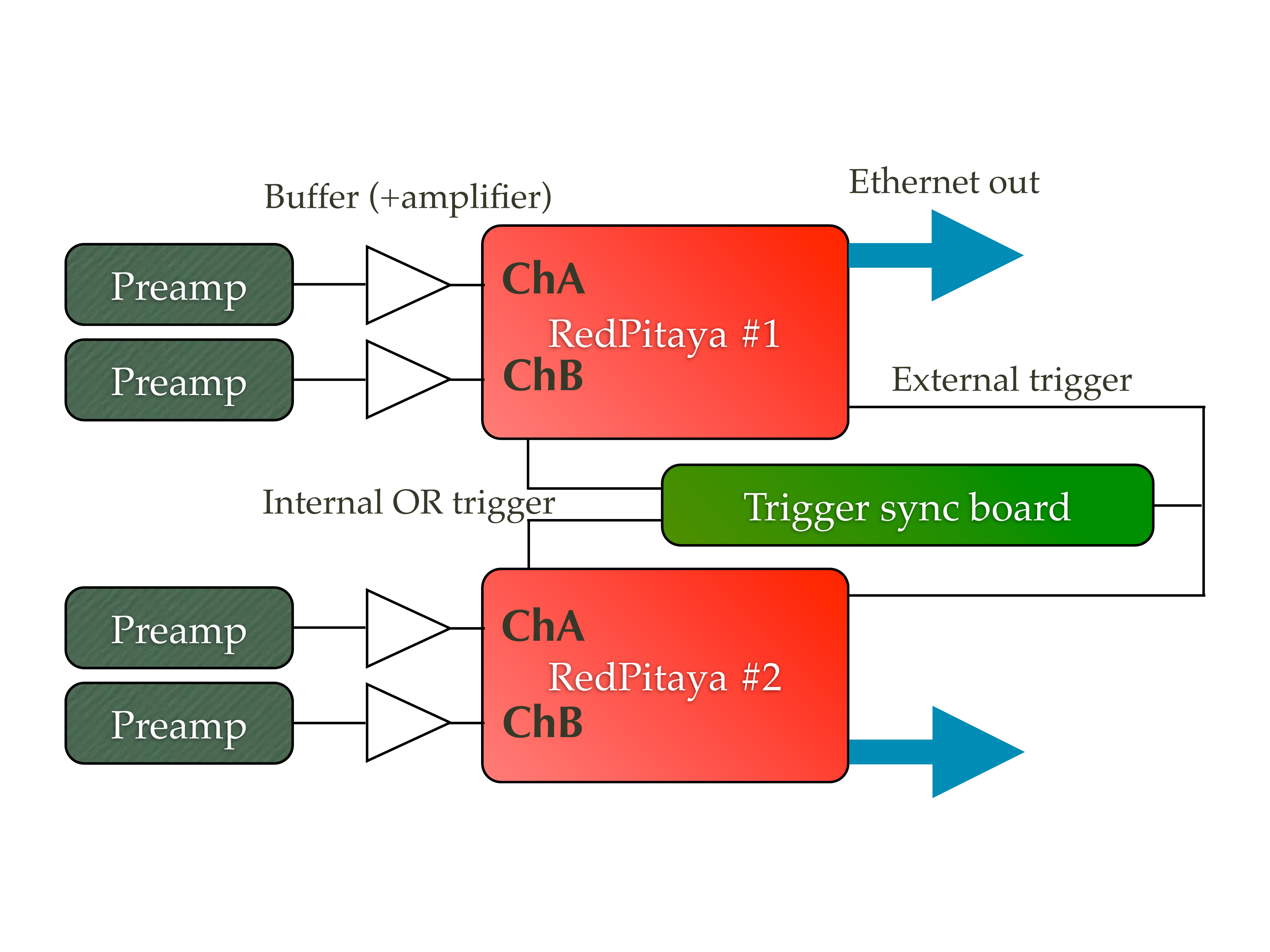}
\caption{\emph{Left panel:} %a first prototype of a 4-channel back-end system, with two RedPitaya boards and the trigger synchronization board. 
the RedPitaya board.
\emph{Right panel:} sketch of the board setup.}
\label{f:pitayas}
\end{figure}

In order to acquire simultaneously the signal from all the bars when one channel is over a given threshold, a modification of the default FPGA configuration has been implemented (Figure~\ref{f:pitayas}, right panel), exploiting the on-board availability of several GPIO pins. A digital signal is sent to an external pin when one or both channels in a board crosses the threshold (internal OR). This signal is mixed in a multi-port digital OR using an external trigger synchronization board, and then feed to the external trigger pin of all boards.
In this way, the effective trigger logic is that of a ``triple OR trigger'' (internal on either channel or external), allowing for the simultaneous acquisition of the waveform across $N$ boards and $2N$ channels.

\begin{figure}[htbp]
\centering
\includegraphics[width=0.98\textwidth]{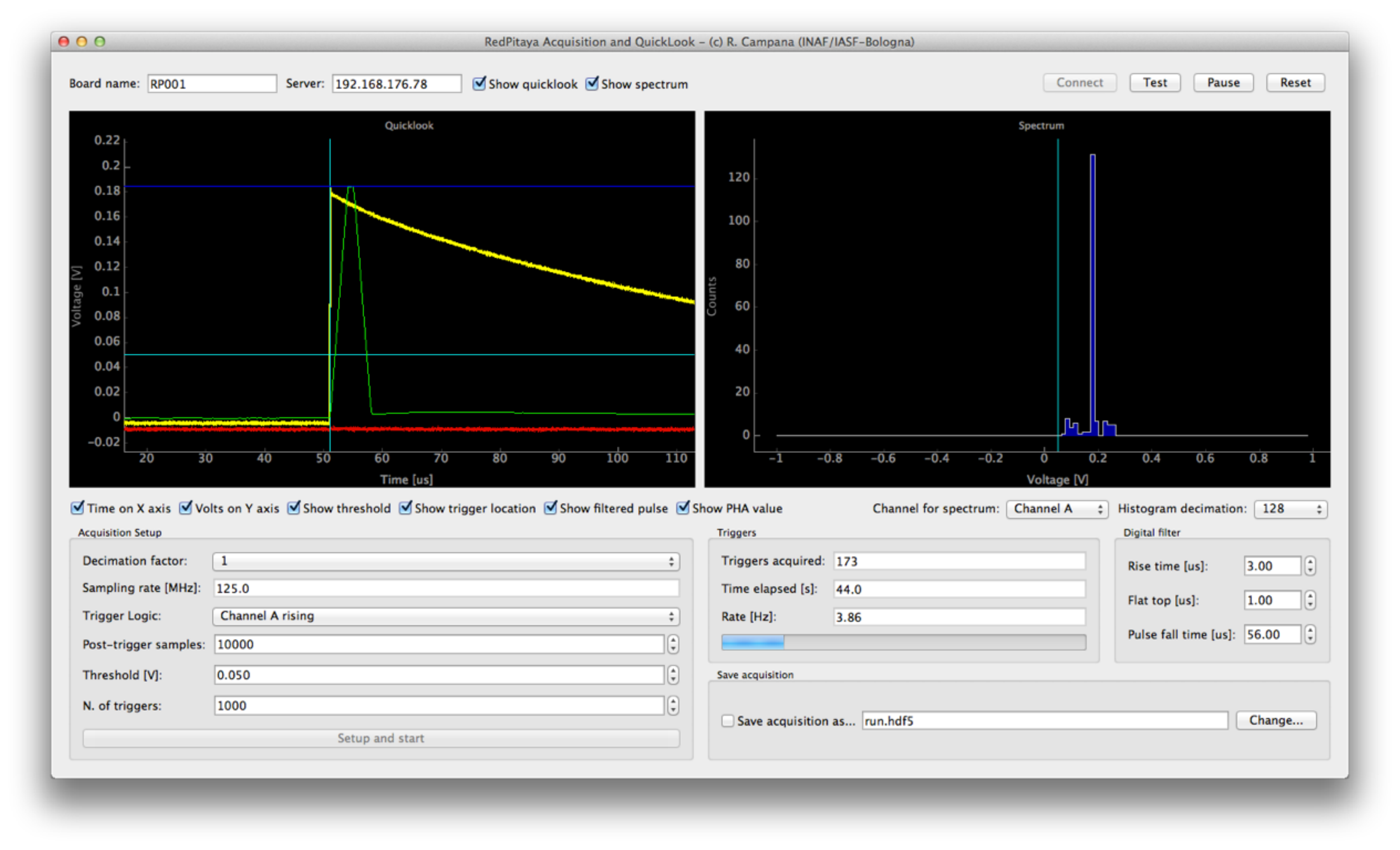}
\caption{Screenshot of the graphical user interface for the acquisition and quicklook of the digital back-end architecture.}
\label{f:pitayagui}
\end{figure}

In Figure~\ref{f:pitayagui} is shown the GUI acquisition and quicklook software, developed at IASF-Bologna in Python using the Qt libraries. 
When launching an acquisition, a server-side program is started on the CPU of each board, configuring the acquisition parameters and the trigger logic and opening a socket for sending the waveforms as Ethernet packets. The client-side program, therefore, acquires these packets and performs digital filtering and spectrum quicklook, beside saving the digitized waveforms in a data file.

\section{CONCLUSIONS}

The preliminary prototypes developed (Section~\ref{s:prototypes}) have shown the good overall performance of the system. We are currently working towards the implementation of the complete XGS prototype, using an innovative, low-cost digital back-end (Section~\ref{s:backend}). The aim is to demonstrate the feasibility of this compact and modular architecture concept (and to drive the development of a dedicated analog front-end electronic circuit) for a space-borne, wide energy range, sensitive photon detector for the future missions for high energy astrophysics.

\acknowledgments % equivalent to \section*{ACKNOWLEDGMENTS}       
 
This project is funded by INAF through a TecnoPRIN 2014 grant.
The Silicon Drift Detectors have been developed in the framework of the ReDSoX collaboration.

% References
\bibliography{bibliography} % bibliography data in report.bib
\bibliographystyle{spiebib} % makes bibtex use spiebib.bst

\end{document}